\documentclass[runningheads]{llncs}
\usepackage{xcolor}
\usepackage{graphicx}
\usepackage{soul}
\setuldepth{Berlin}

\begin{document}
\title{Designing Pids for Reproducible Science Using Time-Series Data}
\author{Wen Ting Maria Tu  \and Stephen Makonin}
\authorrunning{Tu and Makonin}
%
\institute{Simon Fraser University, Canada
}
\maketitle 
\begin{abstract}
As part of the investigation done by the IEEE Standards Association P2957 Working Group, called Big Data Governance and Metadata Management, the use of persistent identifiers (PIDs) is looked at for tackling the problem of reproducible research and science. This short paper proposes a preliminary method using PIDs to reproduce research results using time-series data. Furthermore, we feel it is possible to use the methodology and design for other types of datasets.
\end{abstract}

\section{Introduction}
Reproducible research and science has become an urgent problem as machine learning, and Ai algorithms become more complex (e.g., deep learning). Part of this reproducibility problem is the lack of specificity as to the exact data used for training and testing these learning systems. Part of the investigative work done by IEEE Standards Association P2957 Data Governance and Metadata Management Working Group (BDGMMWG) is to look at how to create a persistent identifier (PID) system that point to specific data within and a number of given and connected datasets.

We first introduce the general architecture proposed by the BDGMMWG\rq{}s Fig. 17 (here as Figure~\ref{fig:bdgmm}) with a discussion on how to use Archival Resource Keys (ARKs) as PIDs in Section~\ref{sec:method}. Following, we describe the design of PIDs for time-series data in Section~\ref{sec:design}. We end with some concluding remarks (Section~\ref{sec:concl}).

\section{Methodology} \label{sec:method}

\begin{figure}
\includegraphics[width=\textwidth]{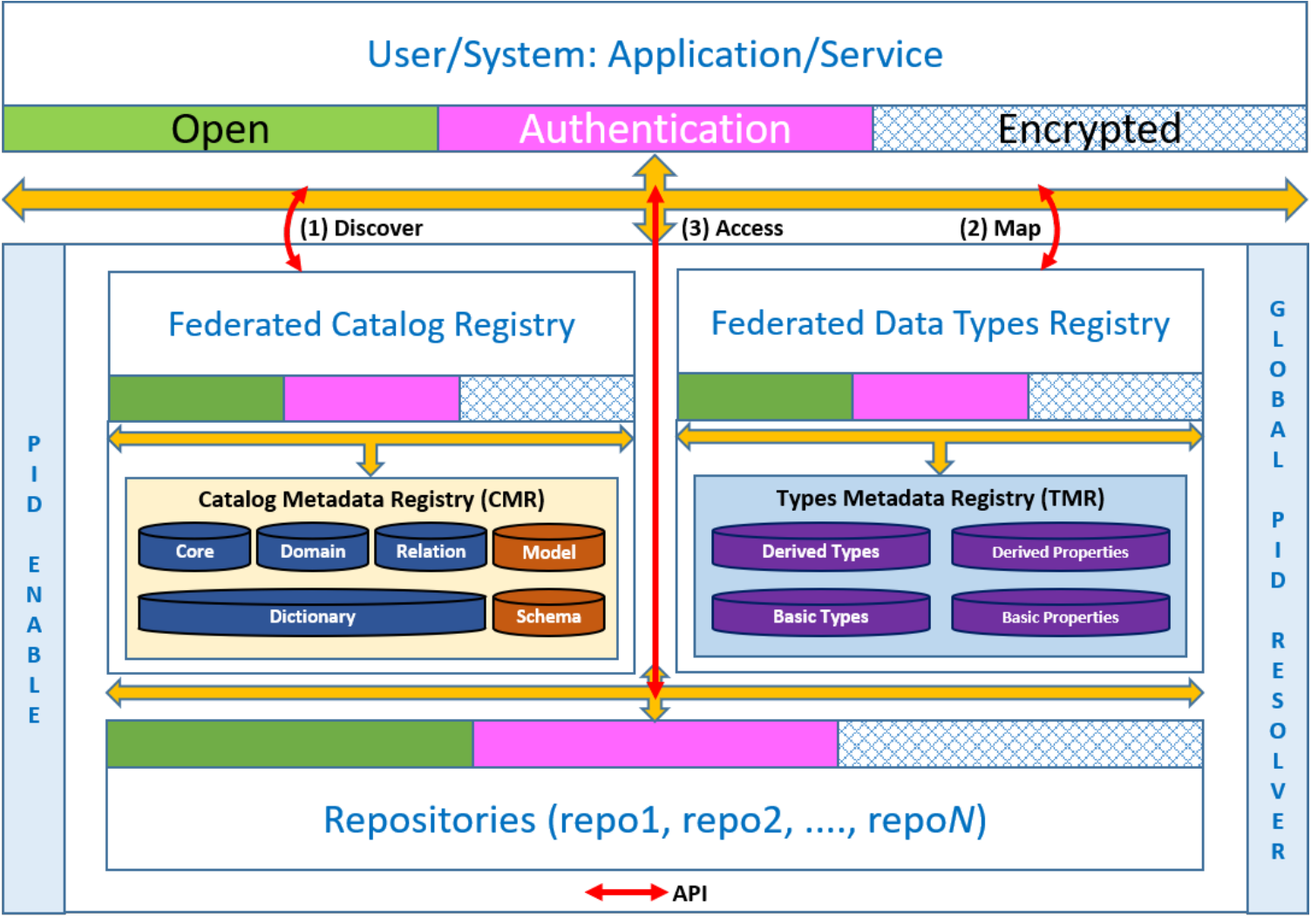}
\vspace{-8mm}
\caption{Block diagram of a federated metadata registry (BDGMMWG Draft Fig. 17~\cite{9133347}).} 
\label{fig:bdgmm}
\vspace{-6mm}
\end{figure}

A federated metadata registry (as in Figure~\ref{fig:bdgmm}) is an extensive database that stores and manages datasets in a centralized location. Within the registry, a federated data catalog system provides a single point of access across all data regardless of the location data is stored. The use of Archival Resource Keys (ARKs) as a type of persistent identifier (PID) to specific data can be queried and accessed from an outside request. Using Nice Opaque Identifier (NOID) within the ARKs PID as a name resolver helps to redirect requests to the proper location of each dataset. We explain the method details in this section.

\subsection{Federated Data Types Registry}

Automated processing of large volumes of data requires implicit details about the data types. Data federation combines all the autonomous data stored to form one extensive database. Data federation stores data into a heterogeneous set. The data is accessible to users as one integrated database using on-demand data integration \cite{van2012data}. A data user using data federation should be able to access different types of database servers and files with various formats from all data sources and be accessible through multiple APIs.

\subsubsection{Types Metadata Registry (TMR)}

A metadata registry is a database used to store, organize, manage, and share metadata \cite{bargmeyer2000metadata}. An issue often encountered in building an extensive metadata registry is that the participants may have used diverse schemas and description methods to create their metadata records \cite{chan2006metadata}. Building one integrated metadata registry enables one search from the user's end to retrieve the information needed rather than searching through all the different existing individual databases. Therefore, interoperability will be a critical component in such development. However, the uniform standardization method is only viable in the early stages of building a metadata registry. Most of the time, different types of resources are already classified under a variety of specialized schemas \cite{chan2006metadata}.

Data Types are characterizations of data at any level of granularity. Data types identified, defined and registered data.  When encountering unknown data, the registry can explicate those types and output the type definitions, relationships and properties. \textit{Basic Types} are used for interpreting and later processing data. A Data Type Registry provides a detailed and structured description of inputted data. Some of the basic types are listed below in Table~\ref{tab:data-types} (left column). \textit{Derived Types} are more specific cases/instances of the basic types. For example, the derived types are listed in Table~\ref{tab:data-types} (right column).

\begin{table}
\centering
\vspace{-5mm}
\caption{Derived Types}\label{tab:data-types}
\begin{tabular}{|l|p{9cm}|}
\hline
Basic Types $\ \ $& Corresponding Derived Types\\
\hline
text        & character, varchar \\ 
number      & integer, long, real, float, double, percentage, scientific \\ 
currency    & USD, RMB, \$, Euro, Yen \\ 
boolean     & check box, yes/no, true/false, on/off\\ 
date/time   & timestamp, short date, medium date, long date, time am/pm, medium time, time 24 hour\\ 
blob        & rich text, attachment, memo, attachment \\
calculated  & lambda unction, imaginary number\\ 
pointer     & hyperlink, lookup \\
\hline
\end{tabular}
\vspace{-6mm}
\end{table}

\textit{Basic Properties} describe the nature of the data. These descriptions are essential to understanding and performing any statistical analysis on the data. Different Exploratory Data Analysis (EDA techniques) can be used to identify the properties of data so that the appropriate statistical methods can be applied to the data. Here is a list of different basic properties of a set of data as shown in Table~\ref{tab:basic-properties} \textit{Derived Properties} are properties that do not exist on the Entity Type associated with the Entity Set; rather, it exists on a type that is derived from the base type of the entity set. 

\begin{table}
\centering
\vspace{-5mm}
\caption{Basic Properties}\label{tab:basic-properties}
\begin{tabular}{|l|}
\hline
Basic Properties\\
\hline
Centre of data\\
Skewness of data\\
Spread among the data members\\
Presence of outliers\\
Correlation among the data\\
Type of probability distribution that the data follows\\
\hline
\end{tabular}
\vspace{-10mm}
\end{table}

\subsection{Federated Catalog Registry}

A data catalog system provides a single place where all available data can be catalogued, enriched, searched, tracked and prioritized. Federation provides a single view across all data of interest to a user, regardless of where the data is stored or sourced.  When there are changes to the external data, the system can quickly crawl external data sources, track changes, make automatic enhancements and push notifications when changes occur. A federated registry is used when user and group information is spread across multiple registries. If the information is stored in various systems, a federated registry can provide a unified view without changing the platform providing a single point of access to reliable data.


\subsubsection{Catalog Metadata Registry (CMR)}

A metadata Catalog System provides a single place where all data can be catalogued, enriched, searched, tracked and prioritized.
Data catalog is a collection of metadata, combined with data management and search tools. The combination serves as an inventory for data and provides tools for data analysis. 
The analyst can search and find data quickly, see all of the available datasets, evaluate  and make informed choices for which data to use, and perform data preparation and analysis efficiently. 

\textit{Core} is the main classifications and groups among the Catalog Metadata Registry. 
\textit{Domain} has three types: environmental domain, object class domain and object format domain.
Environmental domain is the discipline and the community that the scheme services. 
Object class domain is the assembly and grouping of similar objects by type and offer multiple ways to define type. 
Object format domain is the object’s composition and what it is made of \cite{greenberg2005understanding}.
\textit{Relation} defines a related resource. The recommended practice is to identify the related resource by means of a URI. If this is not possible or feasible, a string conforming to a formal identification system may be provided \cite{baker2009guidelines}.
\textit{Model} describes the physical raw data files such as: binary, images, or text containing numeric values\cite{YANG2013612}. 
\textit{Dictionary} is a set of well defined terminology. The goal for a dictionary is to eliminate ambiguity, control the use of synonyms, establish formal relationship among terms and validate terms \cite{doi:10.1080/03615260801973364}. 
\textit{Schema} is also called Formats or Element Sets. It is a set of semantic properties used to describe resources; examples include: MARC 21, Dublin Core, MODS, and ONIX \cite{doi:10.1080/03615260801973364}.

\subsection{ARKs as Persistent Identifiers (PIDs)}

Archival Resource Keys (ARKs) are a form of persistent identifiers (PIDs) with the main focus on web addresses~\cite{arks_2022}. ARKs are open and decentralized persistent identifiers that can be applied to any URL redirects.  To use ARKs, the following items are needed: a Name Assigning Authority Number (NAAN), a minter, a resolver, a plan, and an access persistence policy.

The NAAN indicates the organization that creates the persistent identifier. To eliminates duplicates, the minter tracks strings that have already been assigned to the resources. Once a string is assigned to the resource, the resolver can redirect the persistent ARK identifier to the current access URL where the resource resides. Pre-planning of ARK features such as suffix pass-through or ARK shoulders should be considered during the implementation. If the resource is kept with persistent access, it needs to be securely preserved to prevent accidental loss and deletion~\cite{alliance_2022}. 

Nice Opaque Identifier (NOID) are combinations of numbers and letters to connect to the objects they identify. NOID can be used to form ARKs. In order to prevent duplication, a database will check that no identifier is minted more than once.
The published URL is kept persistent while NOID can be set up as a name resolver to redirect to the new location of where the object resides. N2T.net exists and functions as a shared global resolver~\cite{n2t.net}.

\section{Designing PIDs for Time-Series Data}  \label{sec:design}

The implementation works with both ARKs and n2t.net to deliver the information extraction from the selected database. Files containing time-series data use the timestamp as their primary key, and individual columns of the readings are returned in a CSV file format as a result. AMPds: Almanac of Minutely Power dataset~\cite{6802949} is used for demonstration purposes as it is a well-documented dataset. 

Time-series data present an interesting case where sensor measurements are recorded over time. AMPds contain electricity, water, and natural gas metering data. These meters are sensors that measure different consumption characteristics. For example, each electricity meter (a total of 21) measures 11 different characteristics -- from instantaneous readings like voltage and power to aggregates like energy. In AMPds, each meter is a separate CSV file where columns are measurements and rows are sampled every minute. In each CSV file the timestamp is the unique row ID or primary key. An exmple of a PID might be:

\begin{center}
\ul{https://n2t.net/ark:/57460/AMPds.DWE.V@13332$\sim$13400}
\end{center}

\noindent where 
\ul{https} is a secure HTTP RESTful request, 
\ul{n2t.net} is the global resolver, 
\ul{ark} is means the PID uses ARKs, 
\ul{57460} is the NAAN for our lab, 
\ul{AMPds} is the name of the dataset, 
\ul{DWE} is the meter or sensor,
\ul{V} is the measurement (column), and
\ul{@13332$\sim$13400} is the timestamp range of the readings (rows).

If the user wants to selects all the reading for a given sensor (e.g., DWE or dishwasher meter) for a give measurement (V or voltage), the following URL would be used:

\begin{center}
\ul{https://n2t.net/ark:/57460/AMPds.DWE.V@*}
\end{center}

\noindent or multiple timestamp (i.e., ID) ranges we would use:

\begin{center}
\ul{https://n2t.net/ark:/57460/AMPds.DWE.V@13332$\sim$13400+24300$\sim$25500}
\end{center}

If we wanted to return multiple readings (e.g., V for voltage and I for current) we would use:

\begin{center}
\ul{https://n2t.net/ark:/57460/AMPds.DWE.V+I@*}
\end{center}

\noindent And, for multiple sensors (e.g., HPE, DWE, and WOE; heatpump, dishwasher, and wall oven) we would use:

\begin{center}
\ul{https://n2t.net/ark:/57460/AMPds.HPE+DWE+WOE.V+I@*}
\end{center}

\noindent Lastly, for machine learning, using cross-fold validation is important for training and testing purposes. In these cases would want to exclude a range of timestamps. Here is an example of how to do that:
 
\begin{center}
\ul{https://n2t.net/ark:/57460/AMPds.DWE.V@\_24300$\sim$25500}
\end{center}

\noindent In this case, we are requesting all data except for the data in the inclusive timestamp range of \ul{24300$\sim$25500}. The underscore is used as \textit{exclusion} as the minus sign is not allowed in ARK PIDs. If we were using 10-fold cross-validation, there would be 10 PIDs needed, one for each fold.

%
%
%
%
%
%
%
%
%
%
%
%
%


\section{Conclusions} \label{sec:concl}

Working with a large amount of data requires either centralized or decentralized data storage for access. A Federated metadata registry is an extensive database that stores and manages data in a centralized location. A federated data catalog system provides a single point of access across all data regardless of the location data is stored. A single PID can be used to access specific data from any number of connected datasets. PIDs make accessing data more accessible and provide an easy way to describe the data for easy reproducibility of research and scientific results.

%
%
%
\bibliographystyle{splncs04}
\bibliography{refs}

\end{document}